\documentclass[useAMS,usenatbib]{mn2e}
\usepackage{graphicx}
\usepackage[modulo,switch]{lineno}

\title[Tests of the asymptotic large frequency separation]
  {Tests of the asymptotic large frequency separation of acoustic oscillations in solar-type and red giant stars}
\author[S. Hekker et al.]
  {S.~Hekker$^{1}$, Y. Elsworth$^2$, S. Basu$^3$, A. Mazumdar$^4$, V. Silva Aguirre$^5$, W.J. Chaplin$^2$\\
  $^1$ Astronomical Institute `Anton Pannekoek', University of Amsterdam, Science Park 904, 1098 HX Amsterdam, the Netherlands\\
  $^2$ School of Physics and Astronomy, University of Birmingham, Edgbaston, Birmingham B15 2TT, UK\\
  $^3$ Department of Astronomy, Yale University, P.O. Box 208101, New Haven CT 06520-8101, USA\\
  $^4$ Homi Bhabha Centre for Science Education, TIFR, V. N. Purav Marg, Mankhurd, Mumbai 400088, India\\
  $^5$ Stellar Astrophysics Centre, Department of Physics and Astronomy, Aarhus University, Ny Munkegade 120, DK-8000 Aarhus C, Denmark}

\pagerange{\pageref{firstpage}--\pageref{lastpage}} \pubyear{2011}

\def\LaTeX{L\kern-.36em\raise.3ex\hbox{a}\kern-.15em
    T\kern-.1667em\lower.7ex\hbox{E}\kern-.125emX}

\begin{document}
\linenumbers
\newcommand{\meandnu} {\langle\Delta\nu\rangle}
\label{firstpage}

\maketitle

\begin{abstract}

Asteroseismology, i.e. the study of the internal structures of stars via their global oscillations, is a valuable tool to obtain stellar parameters such as mass, radius, surface gravity and mean density. These parameters can be obtained using certain scaling relations which are based on an asymptotic approximation. Usually the observed oscillation parameters are assumed to follow these scaling relations. Recently, it has been questioned whether this is a valid approach, i.e., whether the order of the observed oscillation modes are high enough to be approximated with an asymptotic theory. In this work we use stellar models to investigate whether the differences between observable oscillation parameters and their asymptotic estimates are indeed significant. We compute the asymptotic values directly from the stellar models and derive the observable values from adiabatic pulsation calculations of the same models.  We find that the extent to which the atmosphere is included in the models is a key parameter. Considering a larger extension of the atmosphere beyond the photosphere reduces the difference between the asymptotic and observable values of the large frequency separation. Therefore, we conclude that the currently suggested discrepancies in the scaling relations might have been overestimated. Hence, based on the results presented here we believe that the suggestions of Mosser et al. (2013) should not be followed without careful consideration.
\end{abstract}


\begin{table*}
\begin{minipage}{\linewidth}
\caption{Summary of the different models used in the current analysis.}
\label{models}
\centering
\renewcommand{\footnoterule}{}
\begin{tabular}{lccccccccc}
\hline\hline
 & EoS & opacity & nuclear reactions & diffusion & atmosphere & X$_0$ & Z$_0$ & mixing length & overshooting\\
\hline
YREC & OPAL \footnote{\citet{rogers2002}, $^b$ \citet{iglesias1996}, $^c$ \citet{adelberger1998}, $^d$ \citet{eddington1926}, $^e$ \citet{ferguson2005} at low $T$, $^f$\citet{formicola2004} for the $^{14}N(p,\gamma)^{15}O$ reaction, $^g$ \citet{angulo1999}, $^h$ \citet{mihalas1988} at low $T$, $^i$ \citet{proffitt1991} for 1.0 and 1..2 M$_{\odot}$ and no diffusion for higher mass models, $^j$ \citet{hopf1934}, $^k$ \citet{krishnaswamy1966}} & OPAL$^b$ & Adelberger$^c$ & no & Eddington$^d$ & 0.707 & 0.019 & 1.8  & 0.2H$_{\rm p}$ \\
 &  & Ferguson$^e$ & Formicola$^f$  &  &  &  &  &  & \\
GARSTEC & OPAL$^a$  & OPAL$^b$ & NACRE$^g$ & no & Eddington$^d$ & 0.701 & 0.019 & solar calibrated & no\\
 & MDH$^g$ & Ferguson$^e$& Formicola$^f$ &  &  &  & & & \\
CESAM & OPAL$^a$ & OPAL$^b$ & NACRE$^g$ & Proffitt / no$^i$ & Hopf$^j$ & 0.699 & 0.020 & 2.0 & no\\
&  & Ferguson$^e$ &  &  &  &  &  &  & \\
MESA & OPAL$^a$  & OPAL$^b$  & NACRE$^g$ & no & Krishna Swamy$^k$ & 0.700 & 0.020  & 1.9 & no\\
 &  & Ferguson$^e$& $^{14}N(p,\gamma)^{15}O$ $^e$ &  &  &  & & & \\
\hline
\end{tabular}
\end{minipage}
\end{table*}

\section{Introduction}
Oscillations in low-mass main-sequence stars, subgiants and red-giant stars are stochastically excited in the turbulent outer layers of these stars. These so-called solar-like oscillations form a distinct pattern of near equidistant peaks in the power spectrum. For p-modes the typical frequency separation between modes of the same degree and consecutive orders is the large frequency separation ($\Delta\nu$). $\Delta\nu$ is related to the sound speed profile in the star and proportional to the square root of the mean density  of the star, i.e.:
\begin{equation}
\Delta\nu=\left (2\int^R_0 \frac{dr}{c} \right ) ^{-1} \propto \sqrt{\overline{\rho}}
\label{dnu}
\end{equation}
with $c$ sound speed, $\overline{\rho}$ mean density and $R$ the stellar radius.

The definition of the large frequency separation relies on the asymptotic theory \citep{tassoul1980,gough1986}. This definition is valid for large values of  the eigenfrequencies corresponding to $\ell / n \rightarrow 0$, with $\ell$ the degree and $n$ the radial order of the mode ($\Delta\nu_{\rm as}$, as defined in Eq.~\ref{dnu}). In practice, the observed large frequency separation ($\Delta\nu_{\rm obs}$) is obtained from the strongest observed oscillation modes, i.e., in the frequency range surrounding $\nu_{\rm max}$ (the frequency of maximum oscillation power). Recently \citet{ mosser2013} (hereafter Mosser13) have presented an investigation regarding the difference between the asymptotic large frequency separation and the large frequency separation observed in low-mass main-sequence stars, subgiants and red-giant stars. To properly account for these differences, Mosser13 derive a relation between the asymptotic and observed large frequency separation. Following this relation they infer $\Delta\nu_{\rm as}$ from $\Delta\nu_{\rm obs}$ for a large sample of stars and suggest new reference values to be used in the asteroseismic scaling relations \citep[e.g.][]{kjeldsen1995}.

The analysis and conclusions presented by Mosser13 are purely based on observed data. Here we present for the first time a relation between the observable and asymptotic large frequency separation based on stellar models. We derive asymptotic values from stellar structure models as well as observable quantities from adiabatic frequency calculations.

\section{Asymptotic relations}
The asymptotic relation that is important here is the equation based on the derivation by \citet{tassoul1980} (see also Mosser13):
\begin{equation}
\nu_{n,\ell}=(n+\frac{\ell}{2}+\epsilon_{\rm as})\Delta\nu_{\rm as}-\frac{\ell(\ell+1)d_0+d_1}{\nu_{n,\ell}}
\label{tassoul}
\end{equation}
with $\nu$ frequency and $\epsilon$ a constant term. The subscript `as' refers to the asymptotic approximation. $d_0$ is related to the gradient of the sound speed integrated over the stellar interior and $d_1$ is a complex function. 

A form of the relation that has common usage in application to observed frequencies is:
\begin{equation}
\nu_{n,\ell}\cong(n+\frac{\ell}{2}+\epsilon_{\rm obs})\Delta\nu_{\rm obs}-\ell(\ell+1)D_0
\label{obs}
\end{equation}
with $D_0 = (\nu_{n,\ell}-\nu_{n-1,\ell+2})/(4\ell+6)$. $\Delta\nu_{\rm obs}$ can be obtained from pair-wise differences of radial frequencies ($\Delta\nu_{\rm obs}=\nu_{n,\ell=0}-\nu_{n-1,\ell=0}$).


Mosser13 state that $\Delta\nu_{\rm obs}$ is not obtained at radial orders that are high enough that $\Delta\nu_{\rm obs}$ is equivalent to $\Delta\nu_{\rm as}$. In this case $\Delta\nu_{\rm obs}$ is not linked to the mean density of the star in the same way as expressed by Eq.~\ref{dnu}.

To investigate this further Mosser13 combined Eqs.~\ref{tassoul}~and~\ref{obs} using only radial modes (Eq.~\ref{eq4}). They included terms to describe the curvature: $A_{\rm as}$ and $\alpha_{\rm obs}$ for the asymptotic and observed curvature, respectively. The curvature accounts for deviations from the regular pattern of radial modes due to stellar internal structure changes. This equation is as follows:
\begin{equation}
\left(n+\epsilon_{\rm as}+\frac{A_{\rm as}}{n}\right)\Delta\nu_{\rm as} = \left(n+\epsilon_{\rm obs}+\frac{\alpha_{\rm obs}}{2}[n-n_{\rm max}]^2\right)\Delta\nu_{\rm obs}
\label{eq4}
\end{equation}
with $n_{\rm max} = \nu_{\rm max} / \Delta \nu_{\rm obs}$ and $\alpha_{\rm obs} =  \rm d \ln \Delta \nu_{\rm obs} / \rm d n$ \citep{mosser2011}.
This results in the following relation between the asymptotic and observed $\Delta\nu$ (Eq. 11 in Mosser13):
\begin{equation}
\Delta\nu_{\rm as}=\Delta\nu_{\rm obs} \left( 1+\frac{n_{\rm max}\alpha_{\rm obs}}{2} \right)
\label{dnu_abs}
\end{equation}
The observed value of $\alpha_{\rm obs}$ is then used to compute the relative difference between $\Delta\nu_{\rm as}$ and $\Delta\nu_{\rm obs}$. This relative difference and thus a probe of $\Delta\nu_{\rm as}$ is in essence computed only from the curvature of the frequencies as opposed to Eq.~\ref{dnu} where $\Delta\nu_{\rm as}$ is computed from an integration of a stellar model.

In this work we derive the relative difference between $\Delta\nu_{\rm as}$ and $\Delta\nu_{\rm obs}$ using stellar models. We use the observationally derived method by Mosser13 (Eq.~\ref{dnu_abs}) on computed adiabatic frequencies as well as $\Delta\nu_{\rm as}$ computed directly from the stellar structure models (Eq.~\ref{dnu}) combined with $\Delta\nu_{\rm obs}$ obtained from adiabatic frequencies.

\section{Stellar evolution models}
In this work we use a large set of YREC models \citep{demarque2008} as well as CESAM \citep{morel1997,morel2008}, GARSTEC \citep{weiss2008} and MESA \citep{paxton2011} models to investigate the model dependence of the results. An overview of the physics included in these models is shown in Table~\ref{models}.

In this work we use sequences of models with masses between 1.0 and 1.6 M$_{\odot}$. These masses  are chosen to bracket the masses of stars presented by Mosser13. The models encompass evolutionary phases from the zero age main sequence up to the tip of the red giant branch (YREC), pre-main-sequence up to the tip of the RGB (GARSTEC), the main-sequence until and including early red-giant phase (CESAM) and main sequence till the end of the He-core-burning phase (MESA).
In addition to the models described above and in Table~\ref{models} we also computed an additional sequence of 1.0 M$_{\odot}$ YREC models constructed with the Krishna Swamy $T$-$\tau$ relation \citep{krishnaswamy1966} and a sequence of 1.0 M$_{\odot}$ YREC models with [Fe/H] = 0.3 dex. Furthermore, there are CESAM models with overshooting of 0.2H$_{\rm p}$ and $Z_0$= 0.01, 0.02, 0.03.

The oscillations for the YREC models are computed using the updated version of the JIG code of \citet{guenther1991}. In a few cases the results were checked using the pulsation code used by \citet{antia1994}. For the GARSTEC, CESAM and MESA models the oscillations are computed using the ADIPLS code \citep{JCD2008}.

\section{Analysis and Results}
We introduce here the analysis and results for the large frequency separations, curvature and resulting differences between $\Delta\nu_{\rm as}$ and $\Delta\nu_{\rm obs}$. We discuss these results in Section 5.
\subsection{Large frequency spacing}
\begin{figure}
\begin{minipage}{\linewidth}
\centering
\includegraphics[width=\linewidth]{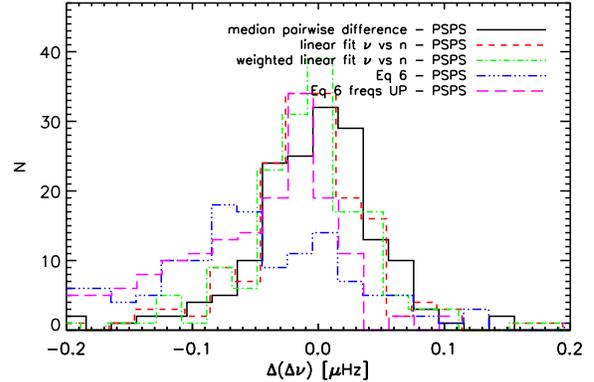}
\end{minipage}
\caption{Histogram of the differences in $\Delta\nu$ from different methods using individual frequencies (see legend and Section 4 for further details) and from the power spectrum of the power spectrum.}
\label{dnutest}
\end{figure}

For the models we compute $\Delta\nu_{\rm obs}$ from the adiabatic frequencies in three different ways. We compute a) the median of the pair-wise differences between radial modes, b)  $\Delta\nu_{\rm obs}$ as the slope of a linear fit  of the frequencies vs. radial order, and c) we use Eq.~10 from Mosser13, namely:
\begin{equation}
\frac{\nu_{n+1,0}-\nu_{n-1,0}}{2}=(1+\alpha_{\rm obs}[n-n_{\rm max}])\Delta\nu_{\rm obs},
\label{eq10M12}
\end{equation}
and apply this to several orders.
The values obtained from these different approaches are consistent within a few percent. Furthermore, we checked how consistent the estimated values of $\Delta\nu_{\rm obs}$ from these methods are with observational techniques such as the power spectrum of the power spectrum (PSPS). A comparison of a subsample of red-giant stars observed with \textit{Kepler} is shown in Fig.~\ref{dnutest}. This figure shows that the median of the pair-wise frequency differences (black solid line) as well as a linear fit of frequencies vs. radial order (red dashed line) provide a $\Delta\nu_{\rm obs}$ value that resembles the one obtained from the PSPS. A Gaussian weighting of the frequencies (green dashed-dotted line) does not have a significant influence on the resulting $\Delta\nu_{\rm obs}$ values (Kolmogorov-Smirnov statistic of 0.056, indicating a significance level of $>99.9$\% for the weighted and unweighted distributions to be similar). $\Delta\nu_{\rm obs}$ obtained from Eq.~\ref{eq10M12} provides less consistent results when used with frequencies measured directly from the highest peaks in the power spectrum (blue dashed-dotted-dotted-dotted line), but this improves when the universal pattern \citep[UP,][]{mosser2011} is used (magenta long dashed line).

\begin{figure}
\begin{minipage}{\linewidth}
\centering
\includegraphics[width=\linewidth]{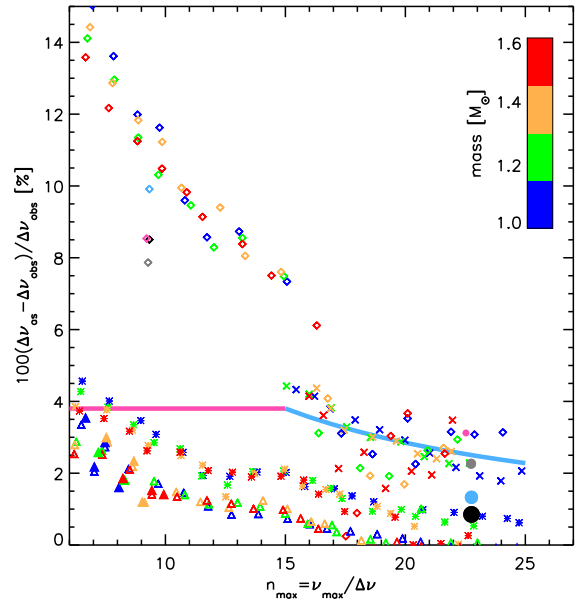}
\end{minipage}
\caption{Relative difference in percent between $\Delta\nu_{\rm as}$ and $\Delta\nu_{\rm obs}$ as a function of $n_{\rm max}$. $\Delta\nu_{\rm obs}$ has been obtained from a linear fit of frequency vs. radial order and $\Delta\nu_{\rm as}$ has been obtained from Eq.~\ref{dnu}. The different colours indicate different stellar masses (see legend). The asterisks indicate YREC models, the diamonds show GARSTEC models, the crosses show CESAM models and the triangles show MESA models. The thick solid lines indicate the fits by Mosser13. Note that for each mass the mean value for each radial order is shown. 
Solar models computed from the YREC code are indicated in black, light blue, gray, and pink and decreasing dot sizes. These indicate a solar model for which the frequency calculation was performed over all grid points, and for models truncated at the surface to all$-$50 grid points, all$-$200 grid points and all$-$500 grid points, respectively.
The sense of stellar evolution is from right to left, except for filled triangles which are in the helium core burning phase.}
\label{reldnu}
\end{figure}

In Fig.~\ref{reldnu} we show the relative difference between $\Delta\nu_{\rm obs}$ and $\Delta\nu_{\rm as}$. These results are based on $\Delta\nu_{\rm as}$ computed following Eq.~\ref{dnu} combined with $\Delta\nu_{\rm obs}$ values computed using a Gaussian weighted linear fit of $\nu$ vs. $n$. To investigate the impact of the frequency range on these results, we computed $\Delta\nu_{\rm obs}$ over 5, 9 and 13 orders centred around $\nu_{\rm max}$ respectively, i.e., $\nu_{\rm max} \pm 2\Delta\nu$,  $\nu_{\rm max} \pm 4\Delta\nu$,  $\nu_{\rm max} \pm 6\Delta\nu$. These showed similar trends and only the results computed over 9 orders are shown in Fig.~\ref{reldnu}. This figure is comparable with figure 6 of Mosser13.

\subsection{Curvature}
\begin{figure*}
\begin{minipage}{0.49\linewidth}
\centering
\includegraphics[width=\linewidth]{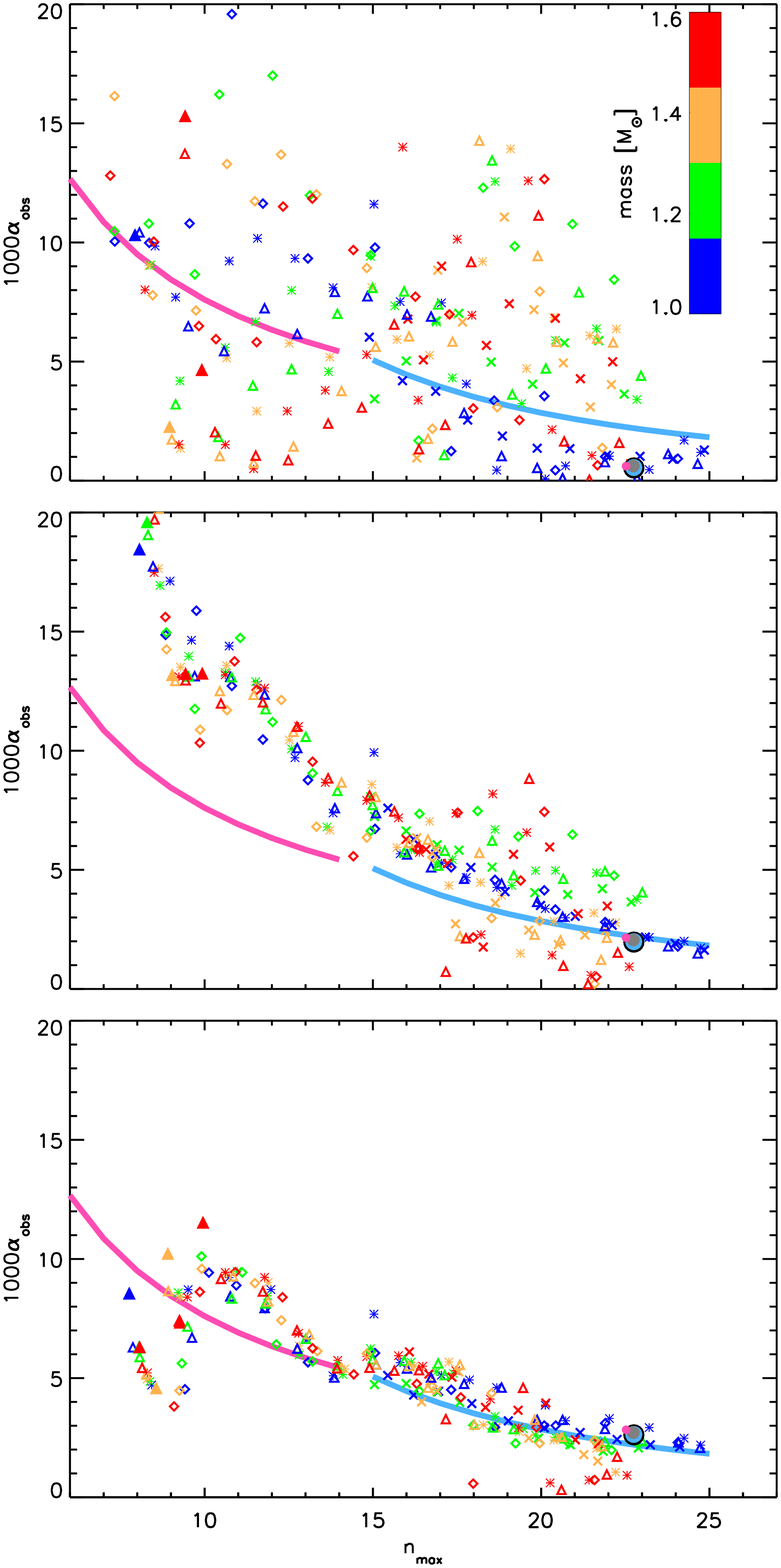}
\end{minipage}
\begin{minipage}{0.49\linewidth}
\centering
\includegraphics[width=\linewidth]{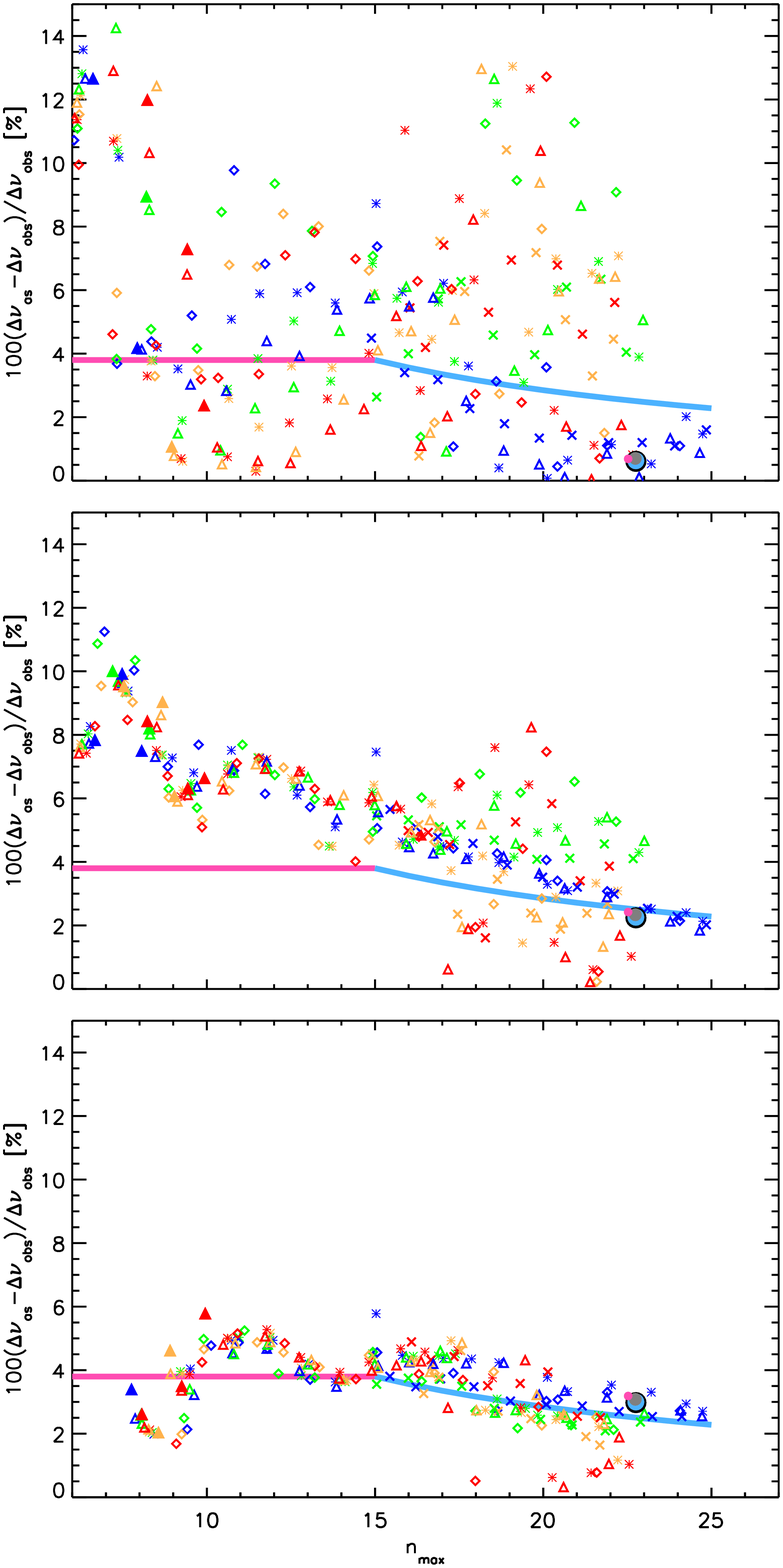}
\end{minipage}
\caption{Left: 1000$\alpha_{\rm obs}$ as a function of $n_{\rm max}$. Right: Relative difference in percent between $\Delta\nu_{\rm as}$ and $\Delta\nu_{\rm obs}$ as a function of $n_{\rm max}$ computed from Eq.~\ref{dnu_abs}.  From top to bottom the frequency range used to determine $\alpha_{\rm obs}$ are 5, 9, 13 orders centred around $\nu_{\rm max}$ respectively. 
In all panels, the colour coding, symbols and thick lines have the same meaning as in Fig.~\ref{reldnu}.  Note that for each mass the mean value for each radial order is shown.} 
\label{alpha}
\end{figure*}

We computed the observed curvature, $\alpha_{\rm obs}$, from the models using $\alpha_{\rm obs} =  \rm d \ln \Delta \nu_{\rm obs} / \rm d n$, i.e., the slope of a linear fit of $\ln \Delta \nu_{\rm obs}$ vs. radial order, and from  Eq.~\ref{eq10M12}. The results of the computations from both formulae are consistent, as they should be, because the formulations are mathematically equivalent.  We show the results in the left panels of Fig.~\ref{alpha}. These graphs are comparable with figure 4 of Mosser13. From top to bottom, we show the impact of the frequency range over which $\alpha_{\rm obs}$ is computed:  5, 9, 13 orders centred around $\nu_{\rm max}$ respectively.

The values of $\alpha_{\rm obs}$ are used to compute the relative difference in $\Delta\nu$ following Eq.~\ref{dnu_abs}. This resembles the approach by Mosser13. The results are shown in the right panels of Fig.~\ref{alpha} as a function of $n_{\rm max}$.



\begin{figure}
\centering
\includegraphics[width=\linewidth]{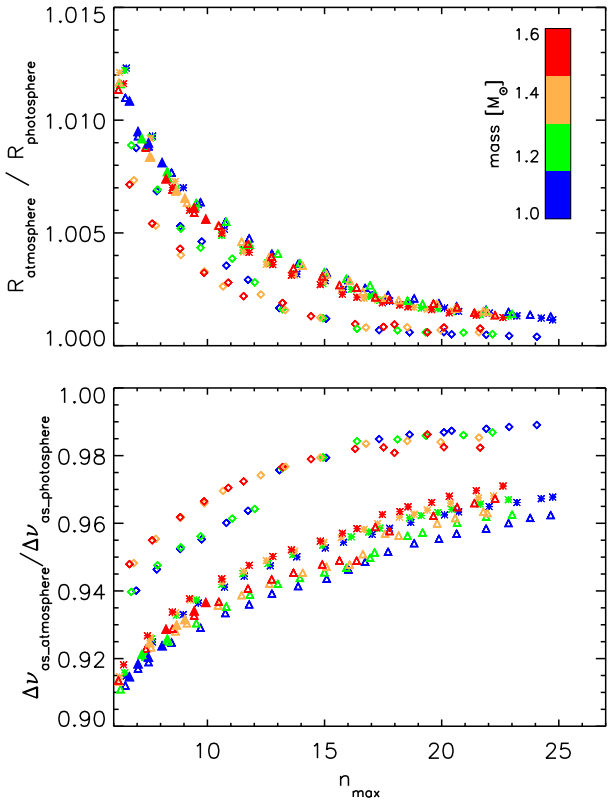}
\caption{Top: the extent of the atmosphere taken into account in the models. Bottom: the resulting $\Delta\nu_{\rm as}$ values normalised by $\Delta\nu_{\rm as}$ at the photosphere. The colours and symbols are the same as in Fig.~\ref{reldnu}.} 
\label{extent}
\end{figure}

\section{Discussion}
In this analysis of stellar models we find that the estimated values of $\alpha_{\rm obs}$ depend on the frequency range over which they are computed. When using 9 orders or wider we obtain for all models a correlation between $\alpha_{\rm obs}$ and $n_{\rm max}$. For 9 orders the curvature obtained from the models exceeds the curvature obtained by Mosser13. This is less so for 13 orders. A similar pattern is present when looking at the relative differences in $\Delta\nu$ computed from $\alpha_{\rm obs}$. 

For the relative difference in $\Delta\nu$ computed using $\Delta\nu_{\rm obs}$ (linear fit of $\nu$ vs. $n$) and $\Delta\nu_{\rm as}$ (Eq.~\ref{dnu}) there is again a correlation with $n_{\rm max}$ (see Fig.~\ref{reldnu}). However, there are clear differences between the results from different models as well as with the results obtained using $\alpha_{\rm obs}$ (Fig.~\ref{alpha}). The reasons for these discrepancies  could possibly lie in the fact that only adiabatic frequencies are used to compute $\alpha_{\rm obs}$ and that a full integration of the stellar models is used in Eq.~\ref{dnu}.

The integration range in Eq.~\ref{dnu} is a possible source of the discrepant results between the different models for the relative difference in $\Delta\nu_{\rm obs}$ - $\Delta\nu_{\rm as}$. For all models the integration is performed over the same radius as used to calculate the frequencies. However the MESA models take a larger radius into account than the YREC models, which in turn take a larger radius into account than the GARSTEC models (see Fig.~\ref{extent}). To verify the impact of the radius that is taken into account, we performed some additional tests for the Sun (dots in Figs~\ref{reldnu} and \ref{alpha}). We use YREC models of the Sun and integrate Eq.~\ref{dnu} over the full radius to obtain $\Delta\nu_{\rm as}$. All YREC models extend to approximately a density of $10^{-10}$, which is equivalent to a maximum fractional radius of 1.0012080 for the solar model, which contains 2721 gridpoints. This full radius is also used to compute the frequencies. Next, we truncate the model by 50, 200 and 500 grid points respectively (equivalent with maximum fractional radii of 1.0010530; 1.0007120; 1.0000990) and compute $\Delta\nu_{\rm as}$ and the frequencies for these truncated models. This cutting off of the atmosphere leads to an increase in the relative difference between $\Delta\nu_{\rm as}$ and $\Delta\nu_{\rm obs}$ when computed using $\Delta\nu_{\rm obs}$ (linear fit of $\nu$ vs. $n$) and $\Delta\nu_{\rm as}$ (Eq.~\ref{dnu}). The results of the truncated models becoming consistent with the results from GARSTEC, CESAM and Mosser13 in Fig.~\ref{reldnu}. By truncating the models the value of $\Delta\nu_{\rm as}$ increases as expected from Eq.~\ref{dnu}, hence increasing the relative difference between the $\Delta\nu_{\rm as}$ and $\Delta\nu_{\rm obs}$ for truncated models of the Sun. 

The effect of the extent of the atmosphere on $\Delta\nu_{\rm as}$ is much larger for giants than for main-sequence stars. This truncation together with the fact that there are differences in the sound speed profiles of the different stellar models most likely cause the larger discrepancies in the present result for giants compared to main-sequence stars. The difference in sound speed profile is most notable for the GARSTEC models for which we find lower values of $\Delta\nu_{\rm as}$ from Eq.~\ref{dnu} than for YREC or MESA models despite the further truncated atmosphere. 

We note that the truncation of the models does not influence the curvature $\alpha_{\rm obs}$ significantly (see Fig.~\ref{alpha}). Furthermore, the tests with different atmospheres and metallicities show that these atmosphere parameters do not have significant influence on the relative difference between $\Delta\nu_{\rm as}$ and $\Delta\nu_{\rm obs}$. However, the use of a free or an isothermal surface boundary condition has influence on the computation of $\Delta\nu_{\rm obs}$ with the free boundary causing an increase in the relative difference in $\Delta\nu$. This effect seems to be secondary to the truncation of the models.

It is not possible to say which of the models represent best the observations in terms of relative difference between $\Delta\nu_{\rm as}$ and $\Delta\nu_{\rm obs}$. What we do know however is that the extent of the atmosphere taken into account does make a difference. For calculating solar frequencies, one obtains a larger surface term correction (a correction needed to account for the fact that the non-adiabatic effects in the outer parts of the stars can not be modelled accurately) for models truncated too close to fractional radius of 1.0. Changing the surface boundary condition is tantamount to changing the surface term. In this light the models that take a larger radius into account could resemble the real relative difference between $\Delta\nu_{\rm as}$ and $\Delta\nu_{\rm obs}$ best.

\section{Conclusions}
For $\alpha_{\rm obs}$ (curvature accounting for deviations from the regular pattern of radial modes due to stellar internal structure changes) obtained over a large frequency range of 13 orders there is remarkable agreement between Mosser13 and the different stellar models. Hence the relative difference between $\Delta\nu_{\rm as}$ and $\Delta\nu_{\rm obs}$ computed from these values of $\alpha_{\rm obs}$ also show general agreement.  We emphasize however that to obtain such reliable values of $\alpha_{\rm obs}$ a large frequency range of preferably 13 orders, i.e., $\nu_{\rm max} \pm 6\Delta\nu$, is needed. For red giants it is typically possible to observe 7 or 8 orders \citep{mosser2010}, which is possibly too limited to obtain a reliable value for $\alpha_{\rm obs}$. For small frequency ranges $\alpha_{\rm obs}$ is poorly defined.

The relative differences between $\Delta\nu_{\rm as}$ and $\Delta\nu_{\rm obs}$ from computations of $\Delta\nu_{\rm as}$ and $\Delta\nu_{\rm obs}$ from stellar models do not show agreement with Mosser13, nor between the different stellar models used. This is mainly due to the extent into the atmosphere used to compute $\Delta\nu_{\rm as}$ and the frequencies. The models including a larger part of the atmosphere in the computation of $\Delta\nu_{\rm as}$ and frequencies show the smallest relative difference between $\Delta\nu_{\rm as}$ and $\Delta\nu_{\rm obs}$. Based on current experience with computation of solar frequencies the inclusion of a larger part of the atmosphere could be best resembling reality. Neither, the atmospheric structure as defined by the $T-\tau$ relations, nor metallicity have significant influence on the relative difference between $\Delta\nu_{\rm as}$ and $\Delta\nu_{\rm obs}$. There are however secondary effects due to the boundary conditions and differences in stellar structure models.

For the models including a larger part of the stellar atmosphere the relative differences between $\Delta\nu_{\rm as}$ and $\Delta\nu_{\rm obs}$ from direct calculations are smaller than the ones observed in Mosser13 indicating that the corrections to the asymptotic scaling relations proposed by Mosser13 might be overestimated. Hence, based on the results presented here we believe that the suggestions of Mosser13 should not be followed without careful consideration. 

\section*{Acknowledgements}
SH acknowledges financial support from the Netherlands Organisation for Scientific Research (NWO). YE and WJC acknowledge support from STFC (The Science and Technology Facilities Council, UK). SB acknowledged NSF grant AST-1105930  and NASA grant NNX13AE70G. AM acknowledges support from the NIUS programme of HBCSE (TIFR). Funding for the Stellar Astrophysics Centre is provided by The Danish National Research Foundation (Grant agreement No. DNRF106). The research is supported by the ASTERISK project (ASTERoseismic Investigations with SONG and {\it Kepler}) funded by the European Research Council (Grant agreement No. 267864). 

\bibliographystyle{mn2e}
\bibliography{asymptotics.bib}

\label{lastpage}

\end{document}